\documentclass[twocolumn]{aastex7}
\received{March 27, 2026}

\graphicspath{{./}{figs/}}

\usepackage{amsmath}
\usepackage{gensymb}
\usepackage{xspace}

\newcommand{\Gaia}{\textit{Gaia}\xspace}

\shorttitle{Locating Blazar Flares with Gaia Astrometry}
\shortauthors{Alexander Plavin}

\begin{document}

\title{\Gaia Sees Blazars Move: Locating Optical Flares Using Astrometry}

\correspondingauthor{A.~V.~Plavin}
\email{alexander@plav.in}

\author[0000-0003-2914-8554]{A.~V.~Plavin}
\email{alexander@plav.in}
\affiliation{Black Hole Initiative, Harvard University, 20 Garden St, Cambridge, MA 02138, USA}
\affiliation{Max-Planck-Institut f\"ur Radioastronomie, Auf dem H\"ugel 69, 53121 Bonn, Germany}

\begin{abstract}
When blazars flare, their optical position moves. We show this by combining \Gaia DR3 proper motions with epoch photometry for blazars with strong optical jet emission. In 60 of 74 sources with significant proper motion, rising flux drives the centroid upstream while fading flux drives it downstream---a near-universal pattern captured by a simple two-component model of constant extended emission and a flaring region. Using this connection, we geometrically localize the optical flares to within $<1$~mas of the VLBI position---a few parsecs at typical blazar distances---placing them in the innermost jet or accretion disk. This purely geometric method requires no multi-wavelength correlations or model-dependent assumptions, and provides an independent spatial anchor for localizing higher-energy flares. Per-epoch astrometry from \Gaia DR4 is set to tighten our constraints even further.
\end{abstract}


\section{Introduction}

Blazars and other jetted active galaxies (AGNs) are among the most variable objects in the sky, with optical flares that can change their brightness by factors of several. This variability spans an enormous range of timescales, from hours to decades; here we focus on the month-scale and longer flares sampled by \Gaia. Even at these timescales, light-travel arguments imply compact emitting regions---but precisely where are these regions located? Near the supermassive black hole, in the accretion disk, or further down the relativistic jet? Answering this question has profound implications for understanding jet launching, particle acceleration, and the connection between accretion and outflow.

The primary method for spatially localizing optical flares has been multi-wavelength time-domain analysis: cross-correlating optical variability with radio flux evolution, with component ejections visibile in radio, or with gamma-ray flares \citep[e.g.,][]{2008Natur.452..966M,2013A&A...552A..11R,2019ApJ...880...32L}. However, this approach has fundamental limitations. It requires assuming that the same physical event is observed across different wavebands---an assumption that fails for ``orphan'' flares appearing in only one band \citep{2019ApJ...880...32L}. Even when correlations exist, interpreting the time lags requires model-dependent assumptions about propagation speeds, and the inferred flare locations can be ambiguous or contradictory---differing not only between sources but between episodes within the same source. A more direct, geometric method for localizing optical flares would be valuable.

Very Long Baseline Interferometry (VLBI) provides milliarcsecond and sub-milliarcsecond astrometry for AGNs at radio wavelengths \citep[for the largest catalog containing tens of thousands of AGNs, see][]{2025ApJS..276...38P}, while \Gaia delivers comparable precision in the optical \citep{2016A&A...595A...1G,2023A&A...674A...1G}. Comparing positions from these two techniques has revealed significant positional offsets for a growing fraction of sources---6\% \citep{2017MNRAS.467L..71P}, 9\% \citep{2019MNRAS.482.3023P}, and 11\% \citep{2022ApJ...939L..32S}---as \Gaia accuracy improved. \citet{2017A&A...598L...1K} discovered that these offsets align with the radio jet direction, a result confirmed by subsequent studies \citep[e.g.,][]{2022A&A...667A.148G}. The dominant explanation is that optical jet emission shifts the \Gaia centroid downstream \citep[e.g.,][]{2019ApJ...871..143P}, although individual sources can involve additional complications \citep{2024A&A...684A.202L,2024evn..conf..141F,2025MNRAS.543..479P,2021A&A...647A.189X}. This jet-driven effect has enabled a range of astrophysical investigations \citep[e.g.,][]{2017MNRAS.471.3775P,2024evn..conf..141F,2024A&A...691A..35B}.

While these positional offsets reflect the time-averaged centroid displacement, they contain no information about how the centroid moves as sources vary. \Gaia proper motions of AGNs have been employed in source selection---to identify candidate doubles, lensed systems, and other unusual objects \citep{2020ApJ...888...73H,2022ApJ...933...28M,2022A&A...660A..16S,2023arXiv230911308K}---but without interpreting the proper motion vectors themselves. \citet{2017MNRAS.471.3775P} predicted that the optical centroid should shift in response to flares, and proposed that such motion could localize the flaring regions. \citet{2019MNRAS.482.3023P} subsequently found that \Gaia proper motions of AGNs preferentially align with the jet direction, confirming that the centroid motion is jet-driven---but did not compare the astrometric signal with optical variability.

Here, we combine \Gaia DR3 proper motions with epoch photometry to show that optical flares universally drive centroid motion in blazars, and use this connection to geometrically localize the flaring regions near the jet base. This study is particularly timely: \Gaia DR3 provides the first combination of proper motions and epoch photometry for a large AGN sample, while DR4---expected in late 2026---will deliver per-epoch astrometry, enabling direct tests of the centroid--flux relationship we predict.

This paper is organized as follows. \autoref{s:data} describes the \Gaia and VLBI data and our sample selection. \autoref{s:scenario} presents the centroid variability model and demonstrates the universal correlation between optical flares and \Gaia proper motion. \autoref{s:localization} develops the \Gaia astrometric model and uses it to constrain the flare position. We summarize our conclusions in \autoref{s:conclusions}.

\section{Data and Sample} \label{s:data}

\subsection{VLBI Positions and Jet Directions}

We obtain VLBI positions from the Radio Fundamental Catalog \citep[RFC;][]{2025ApJS..276...38P}, which contains 21\,949 sources (predominantly, blazars). The catalog is built upon VLBI observations from the 1990s to the present, at frequencies from 1.4 to 86~GHz, with astrometry strongly dominated by 2--8~GHz observations. The RFC provides precise absolute positions at radio wavelengths with typical uncertainties of $\sim$0.4~mas, serving as our astrometric reference frame. These positions correspond to a point close to the jet base, typically within a fraction of a millisecond \citep{2009A&A...505L...1P,2025ApJS..276...38P}.

Jet position angles are taken from \citet{2022ApJS..260....4P}, who determined parsec-scale jet directions for 9220 AGNs using the same underlying VLBI data, at angular scales from sub-mas to $\sim$10~mas from the core, predominantly at 2--15~GHz. The typical accuracy of jet direction measurements is $< 10\degree$.

\subsection{\Gaia DR3 Astrometry and Photometry}

We use astrometric and photometric data from \Gaia Data Release 3 \citep{2023A&A...674A...1G}, which covers the time range from July 2014 to May 2017 ($\Delta T_\mathrm{Gaia} \approx 3$~years). For each source, \Gaia DR3 estimates their time-averaged positions, proper motions, parallaxes and their uncertainties. With a point spread function (PSF) of $\sim$0\farcs2, \Gaia does not resolve the inner structure of a typical AGN, but measures the centroid of all emission falling within the PSF. For extragalactic sources at cosmological distances, true proper motions and parallaxes are negligibly small; non-zero values measured by \Gaia reflect apparent shifts of the optical emission centroid driven by intrinsic source variability \citep{2022A&A...667A.148G}. In addition to astrometric parameters, \Gaia also provides epoch photometry, i.e. light curves. Typically, they contain $\sim$40 data points over $\sim$2.5~yr, very irregularly sampled.

We cross-match RFC and \Gaia by selecting the closest \Gaia counterpart for each RFC source, retaining only pairs whose chance coincidence probability---estimated from the local density of \Gaia sources within $300''$---is below $1/N_\mathrm{RFC}$. This criterion corresponds to fewer than one spurious match expected across the entire catalog. This yields 12\,777 matches, of which 6436 have VLBI jet direction measurements and 4561 have \Gaia epoch photometry. Typical \Gaia uncertainties for our sources are 0.17~mas in position and 0.22~mas\,yr$^{-1}$ in proper motion.

\subsection{Sample Selection}

\begin{figure}
    \centering
    \includegraphics[width=\linewidth]{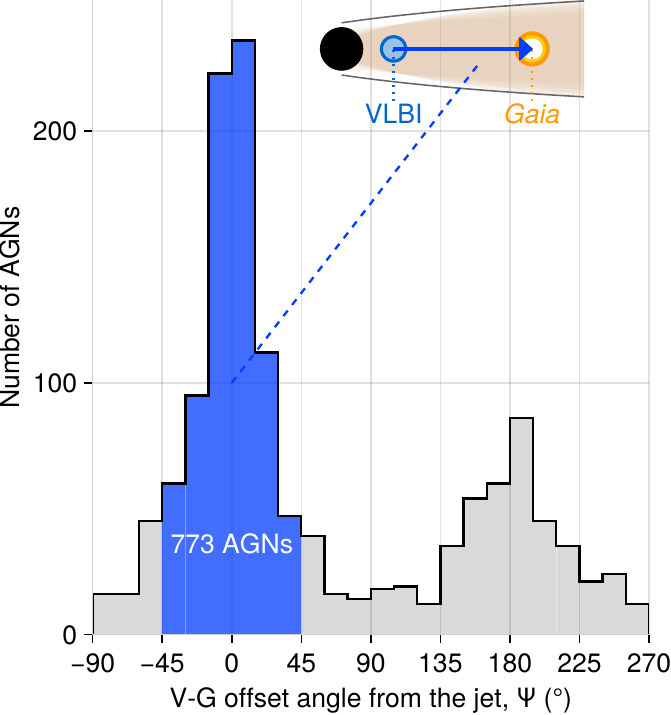}
    \caption{
        Distribution of VLBI-to-\Gaia positional offset directions relative to the parsec-scale jet direction, for sources with statistically significant offsets ($>3\sigma$).
        The peak near $0\degree$ reflects optical jet emission shifting the \Gaia centroid downstream along the jet \citep[see][for a detailed study]{2019ApJ...871..143P}.
        The shaded region highlights the jet-aligned subsample ($<45\degree$) analyzed in this work.\\
        An interactive version of this figure is available at \citet{gaia_agn_motion_code}.
    }
    \label{f:offset_hist}
\end{figure}

Our parent sample consists of AGNs with both \Gaia DR3 astrometry and VLBI positions available. To focus on sources where \Gaia clearly detects optical emission from the extended jet (rather than just the unresolved core), we apply the following selection criteria:
\begin{itemize}
    \item VLBI--\Gaia positional offset significant at $>3\sigma$;
    \item Offset aligned with the VLBI jet direction (within $< 45\degree$), see \autoref{f:offset_hist}
\end{itemize}
These criteria ensure that the \Gaia optical centroid is displaced from the VLBI radio position along the jet, indicating that \Gaia resolves extended jet emission. This selection yields a sample of 773 AGNs.

For most of the analysis, we use only sources with epoch photometry and significantly detected \Gaia proper motions (above $3\sigma$), resulting in a final sample of 74 AGNs, although we perform a cross-check with lower-significance proper motions as well.


\section{Flare-Driven Centroid Motion} \label{s:scenario}

\begin{figure}
    \centering
    \includegraphics[width=\linewidth]{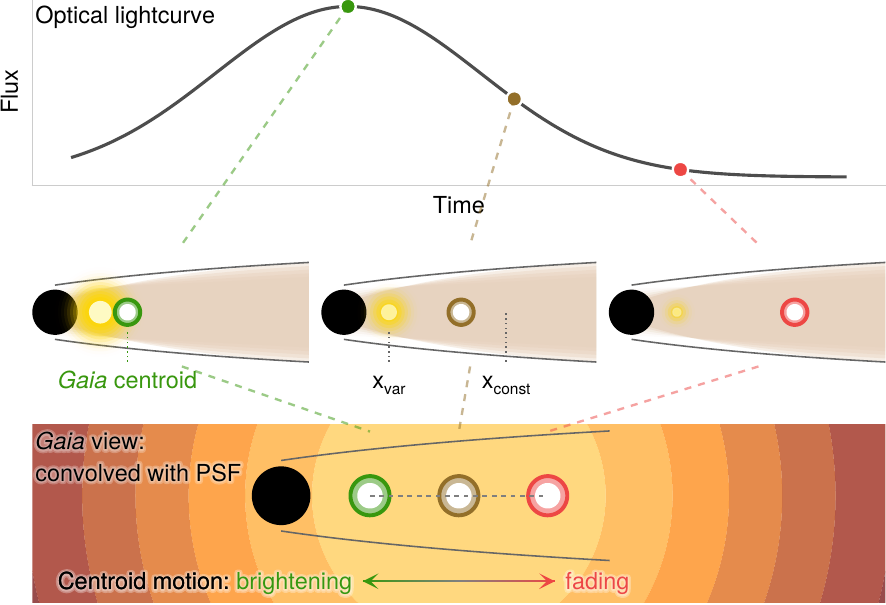}
    \caption{Schematic of how optical flares drive systematic \Gaia position shifts. At peak brightness (green), the centroid is pulled upstream toward the compact flare; as it fades (red), the centroid drifts back downstream toward the extended jet emission. \Gaia cannot resolve these structures---its PSF is far larger than the jet scale (bottom)---but sub-milliarcsecond centroid precision captures the shift.\\ An interactive version of this figure is available at \citet{gaia_agn_motion_code}.}
    \label{f:scenario}
\end{figure}

In the simplest model, the optical emission of an blazar can be decomposed into a constant component and a time-varying one. Given the typical timescales of optical flares (months or shorter; see example \Gaia light curves in \autoref{f:examples}), the varying component should be compact, and here we consider it effectively point-like. The main unknown is where this flaring component is located within the AGN: somewhere in the extended parsec-scale jet? At the effective jet base? In the accretion disk region?

These scales in blazars cannot be directly resolved by optical telescopes. \Gaia provides precise astrometry, tracing the centroid location of the optical emission. Comparing the centroid motion to optical flares can serve as a way to localize the flaring component within the AGN and to compare a range of possible scenarios. These considerations were discussed earlier by \citet{2017MNRAS.471.3775P}, but in this paper we provide the first observational results (using \Gaia DR3 data) and offer predictions for \Gaia DR4, which is expected to contain more detailed astrometric measurements.

\subsection{Two-Component Model} \label{s:centroid}

Quantitatively, the centroid position on the sky can be expressed as follows. Let $S_\mathrm{const}$ and $S_\mathrm{var}(t)$ denote the constant and variable components of the optical flux, respectively, with the total flux given by
\begin{equation}
    S_\mathrm{tot}(t) = S_\mathrm{const} + S_\mathrm{var}(t).
\end{equation}
The position of the varying component is $\mathbf{x}_\mathrm{var}$, and the centroid of the constant component is $\mathbf{x}_\mathrm{const}$ (its brightness profile can be arbitrary). The overall centroid position is then
\begin{equation}
    \mathbf{x}_\mathrm{centroid}(t) = \frac{S_\mathrm{const} \, \mathbf{x}_\mathrm{const} + S_\mathrm{var}(t) \, \mathbf{x}_\mathrm{var}}{S_\mathrm{tot}(t)},
\end{equation}
which can be rewritten as
\begin{equation} \label{eq:centroid}
    \mathbf{x}_\mathrm{centroid}(t) = \mathbf{x}_\mathrm{var} + S_\mathrm{tot}^{-1}(t) \, \mathbf{F}_\mathrm{const},
\end{equation}
where $\mathbf{F}_\mathrm{const} = S_\mathrm{const} \left( \mathbf{x}_\mathrm{const} - \mathbf{x}_\mathrm{var} \right)$ is the constant ``flux--separation product'' vector of the extended emission.

This relation directly connects the centroid motion and the flaring activity. The centroid of the constant emission, $\mathbf{x}_\mathrm{const}$, lies downstream along the jet (it is dominated by the extended jet emission), so $\mathbf{F}_\mathrm{const}$ points downstream for any flaring region located upstream of $\mathbf{x}_\mathrm{const}$ (\autoref{f:scenario}). When the flux rises, $S_\mathrm{tot}^{-1}$ decreases and the centroid shifts toward $\mathbf{x}_\mathrm{var}$; when it fades, the centroid drifts back toward the constant emission. Since $\mathbf{F}_\mathrm{const}$ points downstream for any reasonable flare location, the model makes a robust qualitative prediction: rising flux should drive the centroid upstream and fading flux downstream. This provides a direct observational test of the two-component model.

\subsection{\Gaia Proper Motions Track Flares} \label{s:pm}

\begin{figure}
    \centering
    \includegraphics[width=1\linewidth]{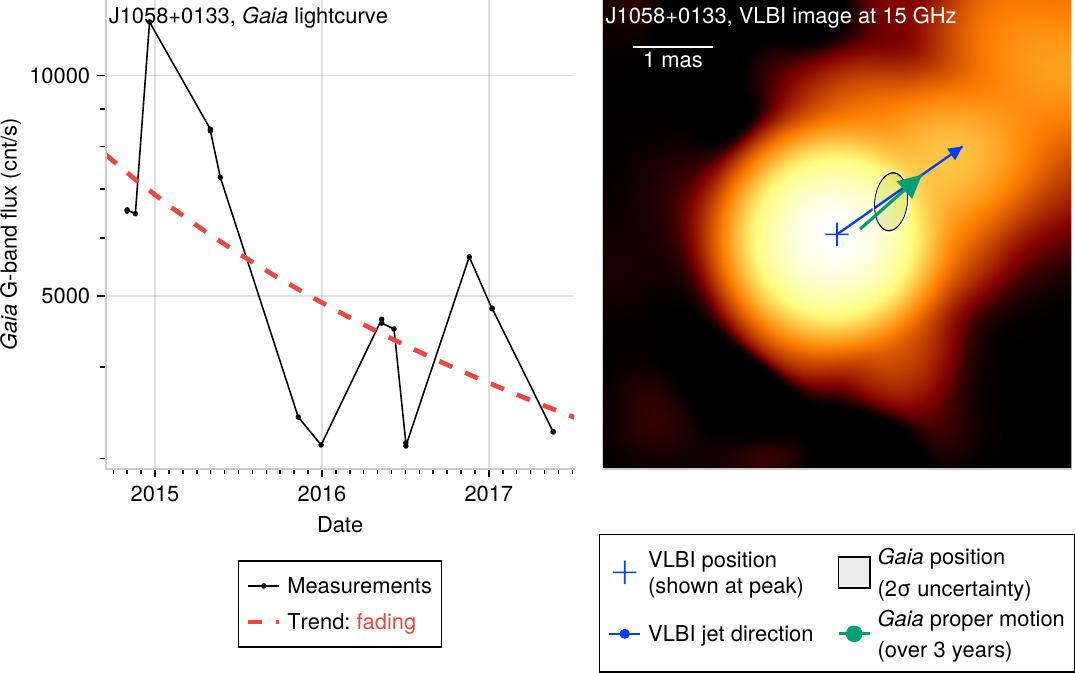}
    \includegraphics[width=1\linewidth]{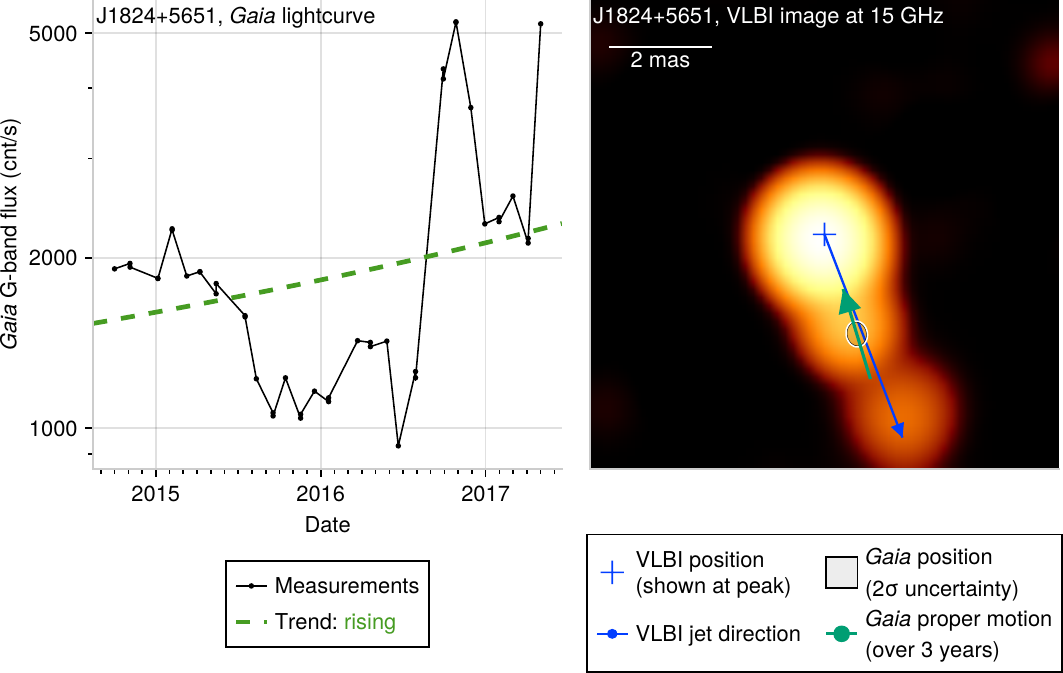}
    \caption{Two example blazars illustrating how flux variability drives centroid motion.\\
    \textit{Top:} J1058+0133 --- fading flux, proper motion downstream along the jet ($0\degree$ in \autoref{f:pm_hist}).\\
    \textit{Bottom:} J1824+5651 --- rising flux, proper motion upstream ($180\degree$ in \autoref{f:pm_hist}). \\
    \textit{Left panels:} \Gaia DR3 light curve with the linear trend in inverse flux (see \autoref{s:centroid}). Colors match those in \autoref{f:pm_hist}.\\
    \textit{Right panels:} VLBI image at 15~GHz, \Gaia position relative to the VLBI position, and \Gaia proper motion vector.}
    \label{f:examples}
\end{figure}

\begin{figure}
    \centering
    \includegraphics[width=\columnwidth]{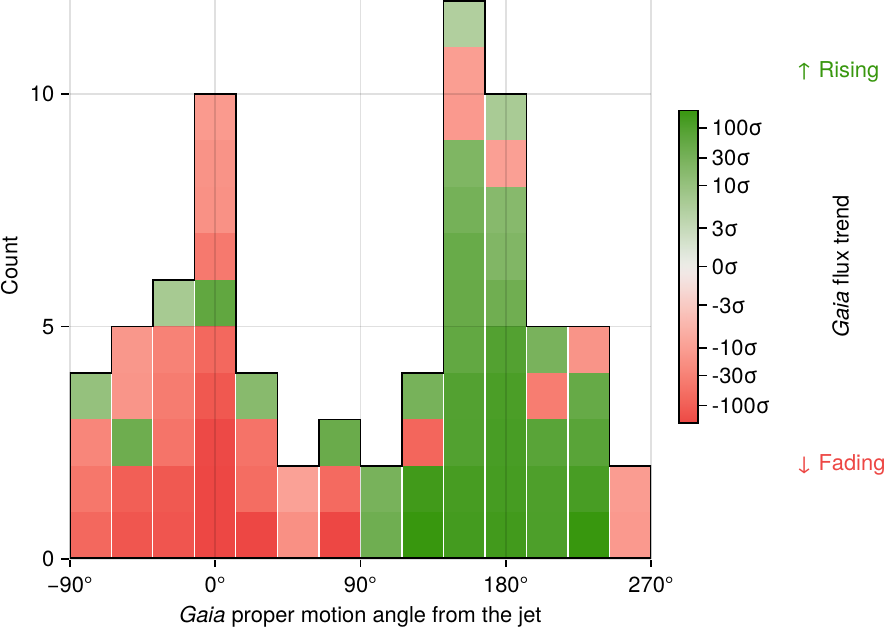}
    \caption{Distribution of \Gaia proper motion directions relative to the jet direction.
    The double-peaked structure shows that proper motions preferentially align with the jet axis.
    Colors encode the \Gaia light curve trend (sign and strength), as in \autoref{f:examples}.
    Nearly all sources follow the same pattern: fading flux drives the centroid downstream ($0\degree$), rising flux drives it upstream ($180\degree$); see \autoref{s:pm} for discussion.\\
    An interactive version of this figure is available at \citet{gaia_agn_motion_code}.}
    \label{f:pm_hist}
\end{figure}

\autoref{f:examples} illustrates the measured centroid motion for two representative AGNs. In J1058+0133, the flux fades and the proper motion points downstream along the jet; in J1824+5651, the flux rises and the proper motion points upstream --- matching the model prediction in both cases.

How common is this pattern? \autoref{f:pm_hist} shows the distribution for all 74 AGNs with significant \Gaia proper motions ($>3\sigma$). The histogram is sharply double-peaked, indicating a strong preference for jet-aligned proper motions. The color coding reveals a near-perfect separation by flux trend: fading sources move downstream, rising sources move upstream. Overall, 60 out of 74 AGNs follow this pattern. Although \autoref{f:pm_hist} displays only AGNs with significantly detected proper motions, we see evidence of the same effect in aggregate for sources with smaller, individually-insignificant proper motions.

The near-universal agreement with the model prediction confirms that the simple two-component picture --- a compact flaring region plus extended constant emission --- captures the dominant mechanism driving \Gaia proper motions in blazars. It also provides a coarse localization: we can already say that the flaring region is upstream of the constant-emission centroid in essentially all sources.

\section{Locating the Flaring Region} \label{s:localization}
\begin{figure*}
    \centering
    \raisebox{-.5\height}{\includegraphics[width=0.35\textwidth]{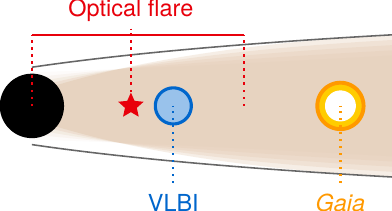}}
    \qquad
    \raisebox{-.5\height}{\includegraphics[width=0.5\textwidth]{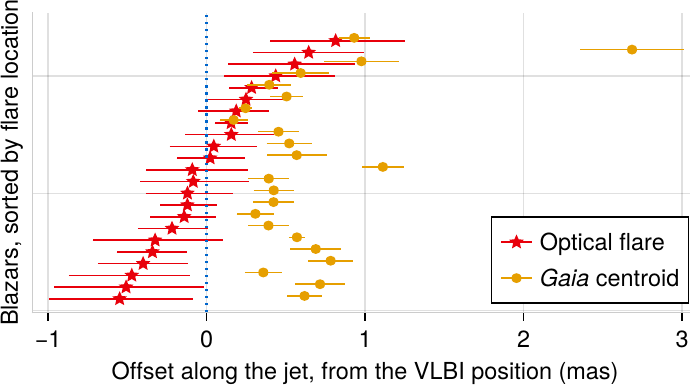}}
    \caption{
        Estimated optical flare position along the jet, relative to the VLBI position, for all AGNs with localization uncertainty below 0.5~mas.\\
        \textit{Left:} schematic of a blazar jet illustrating the relevant components.\\
        \textit{Right:} estimated flare positions derived from the model in \autoref{s:gaiamodel}, and measured \Gaia centroid positions for each source, projected along the jet direction.
        Full image-plane localizations are shown in \autoref{f:flare2d}.
        }
    \label{f:flare_location}
\end{figure*}

Having confirmed that the two-component model correctly predicts \Gaia proper motion directions in AGNs, we now introduce the minimal machinery to localize the flaring region quantitatively.

\subsection{What \Gaia Proper Motions Encode} \label{s:gaiamodel}

In the two-component model (\autoref{eq:centroid}), the centroid position is a linear function of the inverse total flux. If both $\mathbf{x}_\mathrm{centroid}(t)$ and $S_\mathrm{tot}(t)$ were available at individual epochs, one could fit this relation directly. However, \Gaia DR3 does not publish individual measured positions. Instead, \Gaia fits the standard 5-parameter astrometric model to the epoch positions:
\begin{equation} \label{eq:gaiamodel}
    \mathbf{x}(t) = \langle \mathbf{x} \rangle + \dot{\mathbf{x}} \, \Delta t + \varpi \, \mathbf{p}(t),
\end{equation}
where $\dot{\mathbf{x}}$ is the proper motion, $\varpi$ the parallax, $\mathbf{p}(t)$ the parallax factor (determined by the source sky position and the Earth's orbital motion), and $\Delta t = t - t_\mathrm{ref}$. This decomposes the centroid trajectory into temporal basis functions: constant ($1$), linear ($\Delta t$), and parallactic ($\mathbf{p}(t)$).

We only use the mean position and proper motion throughout this paper, ignoring the parallax component. In principle, apparent parallax measurements could improve the identifiability of the model and reduce uncertainties, but we found that including them does not lead to any meaningful changes.

Our model (\autoref{eq:centroid}) is linear in both the unknowns $(\mathbf{x}_\mathrm{var}, \mathbf{F}_\mathrm{const})$ and the observable $S_\mathrm{tot}^{-1}(t)$. The \Gaia astrometric solution is a linear projection of the epoch positions onto temporal basis functions. Composing these two linear operations, the mean position and proper motion can be written as:
\begin{equation} \label{eq:system}
    \left\{\begin{aligned}
    \langle \mathbf{x} \rangle &= \mathbf{x}_\mathrm{var} + \langle S_\mathrm{tot}^{-1} \rangle \, \mathbf{F}_\mathrm{const}, \\
    \dot{\mathbf{x}} &= \dot{S}_\mathrm{tot}^{-1} \, \mathbf{F}_\mathrm{const},
    \end{aligned}\right.
\end{equation}
where $\langle S_\mathrm{tot}^{-1} \rangle$ and $\dot{S}_\mathrm{tot}^{-1}$ are the mean and linear trend of $S_\mathrm{tot}^{-1}(t)$, both computed from the \Gaia light curve. This system is exactly determined and yields the flaring component position:
\begin{equation} \label{eq:rvar}
    \mathbf{x}_\mathrm{var} = \langle \mathbf{x} \rangle
        - \frac{\langle S_\mathrm{tot}^{-1} \rangle}{\dot{S}_\mathrm{tot}^{-1}} \, \dot{\mathbf{x}}.
\end{equation}

This derivation assumes coincident astrometric and photometric epochs and equal weighting of all epochs in the astrometric fit. We believe both approximations are adequate for this pilot analysis; per-epoch astrometry in \Gaia DR4 will remove them entirely (see \autoref{s:dr4} for predictions).

In practice, we express positions relative to the VLBI position, which lies close to the apparent jet base (\autoref{s:data})---making it the natural reference point. With this choice, $\langle \mathbf{x} \rangle$ is the VLBI-to-\Gaia positional offset and $\mathbf{x}_\mathrm{var}$ gives the flare location relative to the VLBI position.

\subsection{Flare Positions}

We solve the system from \autoref{s:gaiamodel} for $\mathbf{x}_\mathrm{var}$ and project along the jet direction ($x = \mathbf{x} \cdot \hat{\mathbf{j}}$, with $\hat{\mathbf{j}}$ pointing downstream along the jet). \autoref{f:flare_location} shows the resulting flare position estimates for the 24 sources with localization uncertainty $\sigma < 0.5$~mas; full image-plane localizations are shown in \autoref{f:flare2d}.

The best-constrained estimates fall within $|x_\mathrm{var}| < 1$~mas of the VLBI position, indicating that the flaring region is located near the jet base. The current localization uncertainties are comparable to or larger than typical radio core-shift magnitudes, making it challenging to pinpoint the flare origin more precisely along the jet. Nevertheless, these results demonstrate the feasibility of localizing optical flares directly from astrometric data, in a purely geometric approach that requires no physical modeling of the emission. Per-epoch astrometry from \Gaia DR4 should substantially sharpen these constraints.

\section{Conclusions} \label{s:conclusions}

We have analyzed the connection between optical variability and astrometric proper motion for AGNs with significant optical emission from the jet---specifically, VLBI-detected sources with \Gaia--VLBI offsets aligned along the jet direction. Our main findings are:

\begin{enumerate}
    \item \textbf{Flares universally drive \Gaia proper motion.} The optical centroid moves upstream during flares (rising flux) and downstream as flares fade, with few exceptions---well captured by a simple two-component model of a flaring core plus constant extended emission.

    \item \textbf{These flares occur near the jet base.} The flaring region is located within $<1$~mas of the VLBI position (a few parsecs at typical blazar distances) in all 24 sources with well-constrained localization, placing it in the innermost jet or accretion disk rather than further downstream in the extended jet.
\end{enumerate}

Together, these results paint a strikingly unified picture of the flaring mechanism for this jet-dominated population, contrasting with the varied and sometimes conflicting results from multi-wavelength time-domain studies. Whether this consistency extends to other AGN populations---such as radio-quiet or disk-dominated sources---remains to be seen.

This study demonstrates that \Gaia astrometry already provides a practical, geometric method for localizing optical flares without requiring multi-wavelength correlations or model-dependent assumptions about propagation speeds. Beyond constraining the optical emission site itself, this provides an independent spatial anchor for localizing higher-energy flares, complementing cross-correlation with VLBI-resolved radio structure. Current precision is limited by the use of time-averaged proper motions rather than per-epoch positions. The upcoming \Gaia DR4 is set to provide per-epoch astrometry, enabling direct measurement of the centroid--flux relationship and substantially tighter localization.

Code to reproduce the analysis and figures from this manuscript is available in \citet{gaia_agn_motion_code}.

\section*{Acknowledgements}

We thank Elena Shablovinskaia
for comments on the manuscript.
A.P.\ is a postdoctoral fellow at the Black Hole Initiative, which is funded by grants from the John Templeton Foundation (grants 60477, 61479, 62286) and the Gordon and Betty Moore Foundation (grant GBMF-8273).
This research was funded in part by the European Union (ERC MuSES project No 101142396).
The views and opinions expressed in this work are those of the authors and do not necessarily reflect the views of these organizations.

This research has made use of data from the Radio Fundamental Catalog \citep{2025ApJS..276...38P}.

\bibliographystyle{aasjournal}
\bibliography{main}

\appendix

\section{Image-Plane Flare Localization} \label{s:flare2d}

\begin{figure*}
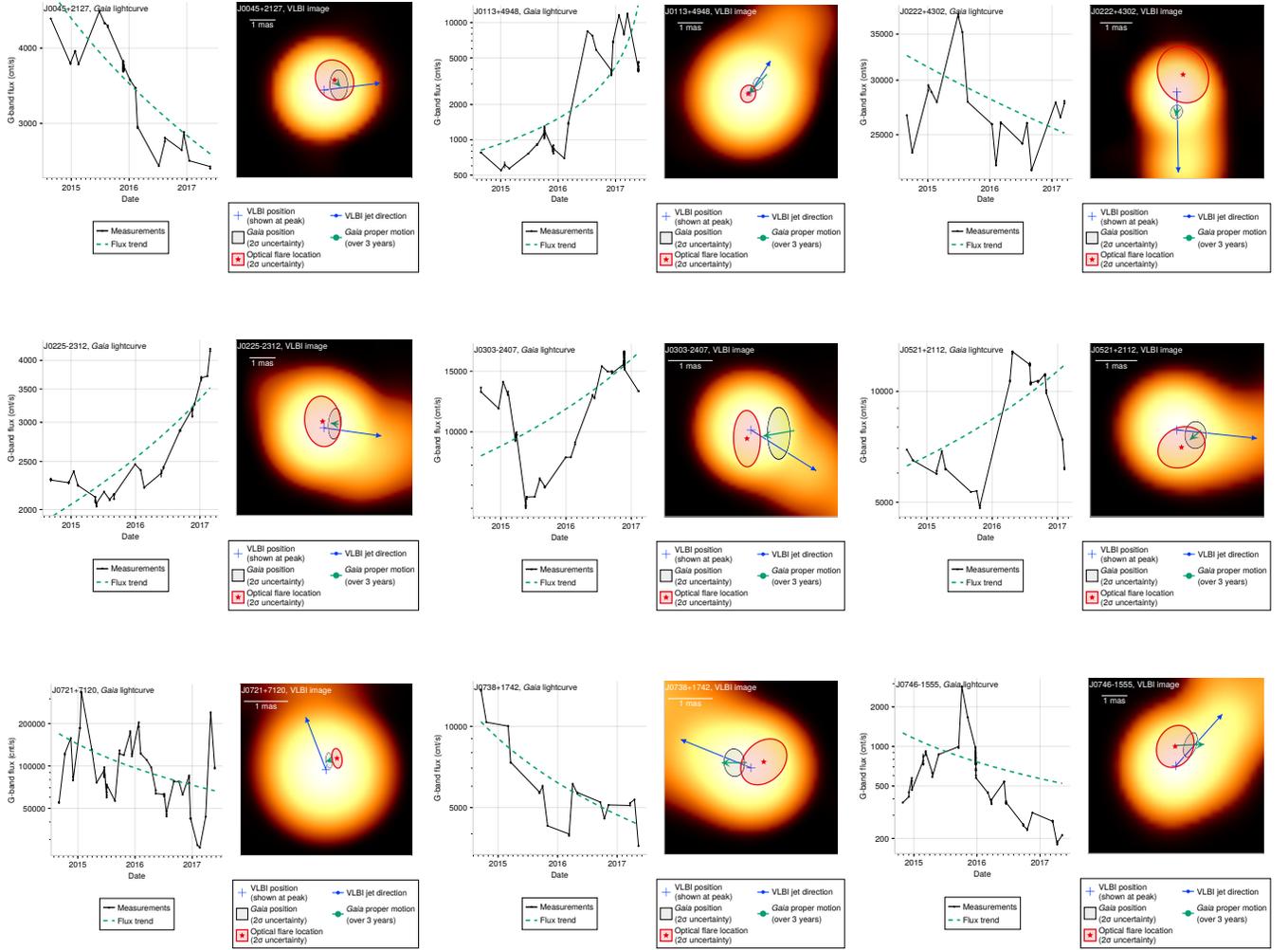

    \centering
    \gridline{
        \fig{flare_2d_clean/J0045+2127.pdf}{0.32\textwidth}{}
        \fig{flare_2d_clean/J0113+4948.pdf}{0.32\textwidth}{}
        \fig{flare_2d_clean/J0222+4302.pdf}{0.32\textwidth}{}
    }
    \gridline{
        \fig{flare_2d_clean/J0225-2312.pdf}{0.32\textwidth}{}
        \fig{flare_2d_clean/J0303-2407.pdf}{0.32\textwidth}{}
        \fig{flare_2d_clean/J0521+2112.pdf}{0.32\textwidth}{}
    }
    \gridline{
        \fig{flare_2d_clean/J0721+7120.pdf}{0.32\textwidth}{}
        \fig{flare_2d_clean/J0738+1742.pdf}{0.32\textwidth}{}
        \fig{flare_2d_clean/J0746-1555.pdf}{0.32\textwidth}{}
    }
    \caption{Image-plane flare localization for all 24 AGNs with flare position uncertainty below 0.5~mas. Each panel shows the \Gaia light curve with flux trend (left) and the VLBI jet image with the \Gaia position, proper motion, and estimated flare location marked (right).}
    \label{f:flare2d}
\end{figure*}

The equations from \autoref{s:gaiamodel} can be applied in the full image plane rather than projected along the jet, yielding direct optical flare localizations for each source. This approach is potentially more sensitive to methodological limitations fundamental at the current stage, particularly those arising from the use of time-aggregated \Gaia astrometric solutions. Still, we find the results informative, and \autoref{f:flare2d} shows the 24 AGNs with flare position uncertainty below 0.5~mas. The inferred flare positions generally coincide with the VLBI position and jet origin. Improved accuracy, both statistical and systematic, is needed to reliably distinguish different localization scenarios.

\begin{figure*}
    \centering
    \gridline{
        \fig{flare_2d_clean/J0757+0956.pdf}{0.32\textwidth}{}
        \fig{flare_2d_clean/J0856-1105.pdf}{0.32\textwidth}{}
        \fig{flare_2d_clean/J0921+6215.pdf}{0.32\textwidth}{}
    }
    \gridline{
        \fig{flare_2d_clean/J1058+0133.pdf}{0.32\textwidth}{}
        \fig{flare_2d_clean/J1221+2813.pdf}{0.32\textwidth}{}
        \fig{flare_2d_clean/J1459+7140.pdf}{0.32\textwidth}{}
    }
    \gridline{
        \fig{flare_2d_clean/J1543+0452.pdf}{0.32\textwidth}{}
        \fig{flare_2d_clean/J1716+6836.pdf}{0.32\textwidth}{}
        \fig{flare_2d_clean/J1728+5013.pdf}{0.32\textwidth}{}
    }
    \gridline{
        \fig{flare_2d_clean/J1734+3857.pdf}{0.32\textwidth}{}
        \fig{flare_2d_clean/J1800+7828.pdf}{0.32\textwidth}{}
        \fig{flare_2d_clean/J1801+4404.pdf}{0.32\textwidth}{}
    }
    \gridline{
        \fig{flare_2d_clean/J1806+6949.pdf}{0.32\textwidth}{}
        \fig{flare_2d_clean/J1927+6117.pdf}{0.32\textwidth}{}
        \fig{flare_2d_clean/J2243-2544.pdf}{0.32\textwidth}{}
    }
    \caption{\autoref{f:flare2d}, continued.}
    \label{f:flare2d2}
\end{figure*}

\clearpage
\section{Predictions for \Gaia DR4} \label{s:dr4}

Throughout this paper, we have relied on time-averaged proper motions and photometric trends to infer the connection between flares and centroid motion. \Gaia DR4 will publish per-epoch astrometry, replacing these aggregated quantities with direct centroid measurements at individual epochs. To illustrate what such data will look like, \autoref{f:dr4pred} shows the centroid position predicted by the two-component model (\autoref{eq:centroid}) at each photometric epoch for the two example sources from \autoref{f:examples}, alongside the linear proper motion from DR3. Where the current analysis sees only a straight line, DR4 should reveal the full centroid trajectory tracing each source's flaring activity. The expected excursions are often on the order of ${\sim}1$~mas, and fitting the model directly to per-epoch data will yield even more precise and assumption-free flare localizations for each AGN.

\begin{figure*}
    \centering
    \gridline{
        \fig{gaia_lc/J1058+0133.pdf}{0.49\linewidth}{}
        \fig{gaia_lc/J1824+5651.pdf}{0.49\linewidth}{}
    }
    \caption{Predicted per-epoch centroid positions for J1058+0133 (\textit{left}) and J1824+5651 (\textit{right}), illustrating the signal expected from \Gaia DR4 per-epoch astrometry.\\
    \textit{Top:} \Gaia light curves analogous to the left panels of \autoref{f:examples}, with the linear trend in inverse flux.\\
    \textit{Bottom panels:} centroid position along the jet predicted at each photometric epoch (black points) compared with the DR3 linear proper motion (blue line). The epoch-by-epoch excursions---driven by flux variability---are the novel observables that DR4 will provide.}
    \label{f:dr4pred}
\end{figure*}

\end{document}